# A modification to Hirsch index allowing comparisons across different scientific fields

Jiří Mazurek

*Department of Informatics and Mathematics,*

*School of Business Administration in Karvina,*

*Karvina, Czech Republic*

mazurek@opf.slu.cz

**Abstract:** The aim of this paper is to propose a simple modification to the original measure, the *relative Hirsch index*, which assigns each researcher a value between 0 (the bottom) and 1 (the top), expressing his/her distance to the top in a given field. By this 'normalization' scholars from different scientific disciplines can be compared.

## 1 Introduction

The Hirsch index (h-index) was introduced by a physicist Jorge E. Hirsch in 2005, originally to determine 'quality' of theoretical physicists by citation counts of their publications. Since then, the h-index was used as a measure of scientific proficiency of scholars in various scientific disciplines, university departments, scientific journals, etc.

However appealing for its simplicity, the h-index has also several drawbacks. Firstly, it does not enable comparisons across different scientific fields due to different citation habits and numbers of researchers active in different fields. Secondly, the h-index does not account for scholars' age, discriminating younger researchers to their older colleagues. Also, h-index cannot distinguish different positions in authors' list of collaborative publications and can be biased by self-citations. Therefore, many modification of h-index were proposed in the last decade, see Alonso et al. (2009), Bornmann (2011), Panaretos and Malesios (2009), Prathap, G. (2006), Lehman et al. (2006) or Zhang 2009.

The aim of this paper is to propose a new simple modification of the original h-index, the *relative Hirsch index* ($h_r$-index) which assigns each researcher a value between 0 (the bottom) and 1 (the top), expressing his/her distance to the top in a given field of science. By this 'normalization' scientist from different disciplines can be compared.

The paper is organized as follows: in the section 2 h-index and its modifications are presented, while section 3 introduces the relative Hirsch index. Conclusions close the article.

## 2 Hirsch index and other citation metrics

Hirsch index assigns each scientist a positive integer number so that a scientist with an index of $h$ has published $h$ papers, and each of them has been cited at least $h$ times, Hirsch (2005). The number of scholars' citation is usually acquired from main bibliographic databases such as ISI Web of Knowledge (WoK), Scopus, Google Scholar or REPEC (for economists). However, the data from these sources differ due to different coverage; see e. g. Bar-Ilan (2007). Moreover, SCOPUS and Google Scholar have limited coverage of publications prior to 1990 according to Meho and Young (2007).

More precise definition of h-index is as follows: Let $f$ be the function assigning each publication $i$ its number of citations, and let the function $f$ be ordered in decreasing order. Then h-index is given as follows:

$$h = \max_i \min(f(i), i) \quad (1)$$

Table 1 provides several indices derived from h-index that avoid some drawbacks mentioned in the previous section.

**Table 1**. Selected citation indices. Source: author

| Index | Explanation |
|---|---|
| c-index | A scientist has c-index $n$ if $n$ of his/her $N$ citations are from authors which are at collaboration distance at least $n$. |
| e-index | The e-index is a complement to the h-index, it takes into account citations beyond $h^2$ core citations, see Zhang (2009). |
| g-index | The g-index can be seen as averaged the h-index |
| m-index | The m-index (m-quotient) is defined as $h/n$, where $n$ is the number of years since the first published paper |
| o-index | The o-index is determined as the geometric mean of the h-index and the most cited paper |
| s-index | The s-index, accounts for the non-entropic distribution of citations, see Silagadze (2009) |
| i10-index | The i10-index provides the number of publications with at least 10 citations |

# 3 The relative Hirsch index

The relative $h_r$-index of a given scientist active in a scientific field *S* is defined as the scientist's *h*-index (1) divided by the current maximal Hirsch index in the field *S*:

$$h_r = \frac{h}{\max_S h} \qquad (2)$$

Clearly, $h_r \in [0,1]$.

The relative Hirsch index expresses scholar's "distance to the top" in his/her field, as $h_r = 1$ is the top (a case of a scientist with the highest h-index in his/her field) and $h_r = 0$ the bottom (a case of a scientist with no citations).

Now consider the following example: Andrei Shleifer is an economist with the highest h-index in his field (*h* = 81, according to WoK). The best physicist is Edward Witten (*h* = 132, WoK). Hence, when comparing a physicist with *h* = 50 ($h_r$ = 0.38), and an economist with *h* = 40 ($h_r$ = 0.49), the economist is performing relatively better than the physicist, because he/she is closer to the top of his/her discipline.

It should be noted that several other papers attempted to overcome a problem with comparisons in different fields, see e.g. Dias (2012). These measures were based on dividing scholars' citations by all citations or by the average number of citation in a given discipline, but these measures were not found suitable. Further, they lack illustrative interpretation, such as a distance to the top, as it is the case of the measure proposed in this paper.

## Conclusions

The aim of the article was to propose the relative Hirsch index which enables comparing scholars in different scientific fields. The index evaluates scholar's distance to the top in his/her discipline. It is a simple measure not accounting for scholar age or a position in authors list, but it can certainly be modified as well. As many other alternatives to original h-index were introduced in the last decade, further research may focus on their (dis)similarity and complementarity.